\documentclass[prl,twocolumn,superscriptaddress,showkeys]{revtex4}
\usepackage{amssymb,epsf,epsfig,amsmath,color}
\usepackage{graphicx}
\usepackage{amssymb}
\usepackage{epstopdf}
\newcommand{\lsim}{
\mathrel{\hbox{\rlap{\hbox{\lower4pt\hbox{$\sim$}}}\hbox{$<$}}}}
\newcommand{\gsim}{
\mathrel{\hbox{\rlap{\hbox{\lower4pt\hbox{$\sim$}}}\hbox{$>$}}}}

\begin{document}
\begin{titlepage}
\vspace*{1.9truecm}

\begin{flushright}
CERN-PH-TH/2008-166\\
SI-HEP-2008-12
\end{flushright}

\vspace*{1.4truecm}

\begin{center}
\boldmath
{\Large{\bf The Golden Modes $B^0\to J/\psi K_{\rm S,L}$ in the \\

\vspace*{0.3truecm}

Era of Precision Flavour Physics}}
\unboldmath
\end{center}

\vspace{1.3truecm}

\begin{center}
{\sc Sven Faller,${}^{a,b}$ Robert Fleischer${}^a$,}
{\sc Martin Jung,${}^{b}$} and {\sc Thomas Mannel${}^{a,b}$}

\vspace{0.5truecm}

${}^a$ {\sl Theory Division, Department of Physics, CERN, CH-1211 Geneva 23,
Switzerland}

\vspace{0.2truecm}

${}^b$ {\sl Theoretische Physik 1, 
Fachbereich Physik, Universit\"at Siegen, D-57068 Siegen, Germany}

\end{center}

\vspace*{1.7cm}

\begin{center}
\large{\bf Abstract}

\vspace*{0.6truecm}

\begin{tabular}{p{14.5truecm}}
{\small The $B^0\to J/\psi K_{\rm S,L}$ channels are outstanding probes of CP 
violation. We have a detailed look at the associated Standard-Model uncertainties, 
which are related to doubly Cabibbo-suppressed penguin contributions, and point
out that these usually neglected effects can actually be taken into account 
unambiguously through the CP asymmetries and the branching ratio of the
$B^0\to J/\psi \pi^0$ decay. Using the most recent $B$-factory measurements, we 
find a negative shift of the extracted value of $\beta$, which softens the tension in 
the fits of the unitarity triangle. In addition, this strategy can be used to constrain a 
possible new-physics phase in $B^0$--$\bar B^0$ mixing. The proposed strategy is 
crucial to fully exploit the tremendous accuracies for the search for this kind of new 
physics that can be achieved at the LHC and future super-flavour factories.}
\end{tabular}

\end{center}

\vspace*{1.7truecm}

\vfill

\noindent
September 2008

\end{titlepage}

\newpage
\thispagestyle{empty}
\mbox{}

\newpage
\thispagestyle{empty}
\mbox{}

\rule{0cm}{23cm}

\newpage
\thispagestyle{empty}
\mbox{}

\setcounter{page}{0}

\preprint{CERN-PH-TH/2008-nnn}

\date{September 4, 2008}

\title{\boldmath The Golden Modes $B^0\to J/\psi K_{\rm S,L}$ in the Era
of Precision Flavour Physics\unboldmath}

\author{Sven Faller}
\affiliation{Theory Division, Department of Physics, CERN, CH-1211 Geneva 23,
Switzerland}
\affiliation{Theoretische Physik 1, Fachbereich Physik, Universit\"at Siegen, 
D-57068 Siegen, Germany}
\author{Martin Jung}
\affiliation{Theoretische Physik 1, Fachbereich Physik, Universit\"at Siegen, 
D-57068 Siegen, Germany}
\author{Robert Fleischer}
\affiliation{Theory Division, Department of Physics, CERN, CH-1211 Geneva 23,
Switzerland}

\author{Thomas Mannel}
\affiliation{Theory Division, Department of Physics, CERN, CH-1211 Geneva 23,
Switzerland}
\affiliation{Theoretische Physik 1, Fachbereich Physik, Universit\"at Siegen, 
D-57068 Siegen, Germany}

\begin{abstract}
\vspace{0.2cm}\noindent
The $B^0\to J/\psi K_{\rm S,L}$ channels are outstanding probes of CP 
violation. We have a detailed look at the associated Standard-Model uncertainties, 
which are related to doubly Cabibbo-suppressed penguin contributions, and point
out that these usually neglected effects can actually be taken into account 
unambiguously through the CP asymmetries and the branching ratio of the
$B^0\to J/\psi \pi^0$ decay. Using the most recent $B$-factory measurements, we 
find a negative shift of the extracted value of $\beta$, which softens the tension in 
the fits of the unitarity triangle. In addition, this strategy can be used to constrain a 
possible new-physics phase in $B^0$--$\bar B^0$ mixing. The proposed strategy is 
crucial to fully exploit the tremendous accuracies for the search for this kind of new 
physics that can be achieved at the LHC and future super-flavour factories.
\end{abstract}

\keywords{CP violation, non-leptonic $B$ decays} 

\maketitle
\noindent
CP-violating effects in $B^0$ decays into CP eigenstates $f$ are 
studied through time-dependent rate asymmetries:
\begin{eqnarray}
\lefteqn{A_{\rm CP}(t;f)\equiv\frac{\Gamma(B^0(t)\to f)-
\Gamma(\bar B^0(t)\to f)}{\Gamma(B^0(t)\to f)+
\Gamma(\bar B^0(t)\to f)}}\nonumber\\
&&=C(f)\cos(\Delta M_dt)-S(f)\sin(\Delta M_dt),\label{CP-asym}
\end{eqnarray}
where $C(f)$ and $S(f)$ describe direct and mixing-induced CP violation, 
respectively. The key application is given by $B^0\to J/\psi K_{\rm S,L}$ decays,
which arise from $\bar b\to \bar c c \bar s$ processes.
If we assume the Standard Model (SM) and neglect doubly Cabibbo-suppressed  
contributions to the $B^0\to J/\psi K^0$ amplitude, we 
obtain \cite{bisa}
\begin{equation}\label{CS-def}
C(J/\psi K_{\rm S,L})\approx 0, \quad 
S(J/\psi K_{\rm S,L})\approx-\eta_{\rm S,L}\sin2\beta, 
\end{equation}
where $\eta_{\rm S}=-1$ and $\eta_{\rm L}=+1$ are the CP eigenvalues 
of the final states, and $\beta$ is an angle of the unitarity triangle 
(UT) of the Cabibbo--Kobayashi--Maskawa (CKM) matrix. The 
usual experimental analyses assume that (\ref{CS-def}) is valid exactly; 
the most recent data then result in
\begin{equation}\label{s2b-psiK}
(\sin 2\beta)_{J/\psi K^0}=0.657 \pm 0.024,
\end{equation} 
which is obtained from the average of the measured $S(J/\psi K_{\rm S,L})$ 
values \cite{BellePsiK,BaBarPsiK}. It is the purpose of the present letter to critically review 
this assumption.

\begin{figure}
    \centering
    \includegraphics[width=6.0truecm]{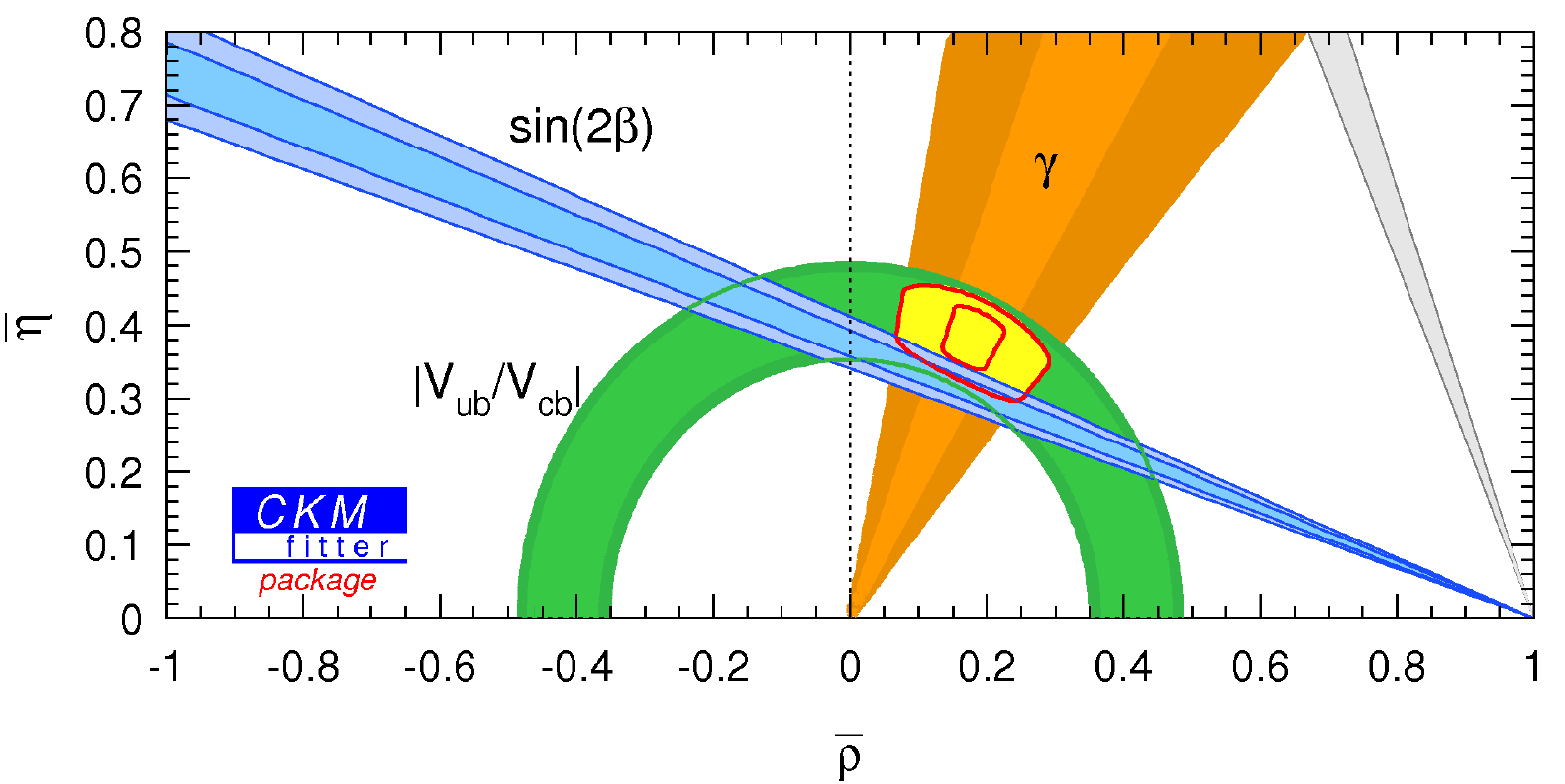} 
      \vspace*{-0.3truecm}
    \caption{Constraints in the $\bar\rho$--$\bar\eta$ plane (1 and $2\,\sigma$ ranges).}\label{fig:1}
 \end{figure}

Using also data for CP violation in $B^0\to J/\psi K^*$ decays \cite{HFAG}, $\beta$ 
can be fixed unambiguously, where the value in (\ref{s2b-psiK}) corresponds 
to $\beta=(20.5\pm0.9)^\circ$.
 In Fig.~\ref{fig:1}, created with the \emph{CKMfitter} software \cite{CKMfitterSoftware},
we show the resulting 
constraint for the apex of the UT in the $\bar\rho$--$\bar\eta$ plane
of the generalized Wolfenstein parameters \cite{wolf,blo}. Moreover, we
include the circle coming from the UT side $R_b\equiv (1-\lambda^2/2)
|V_{ub}/(\lambda V_{cb})|$, where 
$\lambda\equiv|V_{us}|=0.22521\pm0.00083$ \cite{CKMfitter}; taking the 
most recent developments in the determination of $|V_{ub}|$ and $|V_{cb}|$ 
from semileptonic $B$ decays into account \cite{PDG}, we find 
$R_b= 0.423 ^{+0.015}_{-0.022}\pm 0.029$, where here and in the following the first
error comes from experiment and the second from theory. We show also the range corresponding 
to $\gamma=(65\pm10)^\circ$, which is well in accordance with
the analyses of the UT in Refs.~\cite{UTfit,CKMfitter} and the information 
from $B_{d,s}\to\pi\pi,\pi K,KK$ decays \cite{RF-BsKK-07}. 
This angle will be determined with 
only a few degrees uncertainty thanks to CP violation measurements in 
pure tree decays at  LHCb (CERN). In analogy to $R_b$, the value of $\gamma$
extracted in this way is expected to be very robust with respect to new-physics 
(NP) effects. In Fig.~\ref{fig:1}, we can see the tension that is also present in more 
refined fits of the UT for a couple of years  \cite{UTfit,CKMfitter}.

Since $B^0$--$\bar B^0$ mixing is a sensitive probe for NP
(see, e.g., \cite{FM-BpsiK,UTfit-NP,BaFl}), this effect could 
be a footprint of such contributions. Provided they are
CP-violating, we have
\begin{equation}
\phi_d=2\beta+\phi_d^{\rm NP},
\end{equation}
where $\phi_d$ denotes the $B^0$--$\bar B^0$ mixing phase
and $\phi_d^{\rm NP}$ is its NP component. If we assume that NP 
has a minor impact on the $B^0\to J/\psi K^0$ amplitude, the relations
in (\ref{CS-def}) remain valid, with the replacement $2\beta \to \phi_d$.

Using Fig.~\ref{fig:1}, the ``true" value of $\beta$ 
can be determined through $R_b$ and tree-level extractions of $\gamma$. 
We find $\beta_{\rm true}= (24.9^{+1.0}_{-1.5} \pm 1.9)^\circ$,
which is essentially 
independent of the error on $\gamma$ for a central value around 
$65^\circ$ (and yields 
$(\sin 2\beta)_{\rm true}=0.76^{+0.02}_{-0.04}{}^{+0.04}_{-0.05}$). 
Consequently,
\begin{equation}\label{2b-diff}
(\phi_d)_{J/\psi K^0}-2\beta_{\rm true}=-(8.7 ^{+2.6}_{-3.6}\pm 3.8)^\circ.
\end{equation}

Let us now have a critical look at the hadronic SM uncertainties affecting the 
extraction of $\phi_d$ from $B^0\to J/\psi K_{\rm S,L}$. In the SM, we may 
write \cite{RF-BpsiK}
\begin{equation}\label{ampl}
A(B^0\to J/\psi K^0)=\left(1- \lambda^2/2\right)
{\cal A}\left[1+\epsilon a e^{i\theta}e^{i\gamma}\right],
\end{equation}
where 
\begin{equation}\label{Ampl-def}
{\cal A}\equiv \lambda^2 A \left[A_{{\rm T}}^{(c)}+A_{{\rm P}}^{(c)}-
A_{{\rm P}}^{(t)}\right]
\end{equation}
and
\begin{equation}\label{a-def}
a e^{i\theta}\equiv R_b
\left[\frac{A_{{\rm P}}^{(u)}-A_{{\rm P}}^{(t)}}{A_{{\rm T}}^{(c)}+
A_{{\rm P}}^{(c)}-A_{{\rm P}}^{(t)}}\right]
\end{equation}
are CP-conserving parameters, with $A_{{\rm T}}^{(c)}$ and $A_{{\rm P}}^{(j)}$ 
denoting strong amplitudes that are related to tree-diagram-like and penguin 
topologies (with internal $j\in\{u,c,t\}$ quarks), respectively, while 
$A\equiv |V_{cb}|/\lambda^2=0.809\pm0.026$ and 
$\epsilon\equiv\lambda^2/(1-\lambda^2)=0.053$ are CKM factors.

Looking at (\ref{ampl}), we observe that $a e^{i\theta}$ enters with the 
tiny parameter $\epsilon$. Therefore, this term is usually neglected, 
which yields (\ref{CS-def}). However, $a e^{i\theta}$ suffers from large 
hadronic uncertainties, and may be enhanced through long-distance
effects.  As discussed in detail in Ref.~\cite{FFM-long}, the generalization 
of these expressions to take also the penguin effects into account reads as follows:
\begin{equation}\label{S-gen}
\frac{-\eta_{\rm S,L}S(J/\psi K_{\rm S,L})}{\sqrt{1-C(J/\psi K_{\rm S,L})^2}}
=\sin(\phi_d+\Delta\phi_d),
\end{equation}
where
 \begin{eqnarray}
\sin\Delta\phi_d&=&\frac{2 \epsilon a\cos\theta \sin\gamma+\epsilon^2a^2
\sin2\gamma}{N\sqrt{1-C(J/\psi K_{\rm S,L})^2}}\label{sDelPhi}\\
\cos\Delta\phi_d&=&\frac{1+ 2 \epsilon a\cos\theta \cos\gamma+\epsilon^2a^2
\cos2\gamma}{N\sqrt{1-C(J/\psi K_{\rm S,L})^2}}\label{cDelPhi}
\end{eqnarray}
with $N\equiv1+2\epsilon a \cos\theta \cos\gamma+\epsilon^2a^2$,
so that
\begin{equation}\label{tDelPhi}
\tan\Delta\phi_d=\frac{2 \epsilon a\cos\theta\sin\gamma+\epsilon^2a^2
\sin2\gamma}{1+ 2 \epsilon a\cos\theta\cos\gamma+\epsilon^2a^2\cos2\gamma}.
\end{equation}
Concerning direct CP violation, we have
\begin{equation}\label{C-psiK}
C(J/\psi K^0)=-0.003 \pm 0.019,
\end{equation} 
which is again an average over the $J/\psi K_{\rm S}$ and $J/\psi K_{\rm L}$ final 
states \cite{BellePsiK,BaBarPsiK}. Consequently, the deviation of the terms 
$\sqrt{1-C(J/\psi K_{\rm S,L})^2}$ from one is at most at the level of $0.0002$, 
and is hence completely negligible. 

In order to probe the importance of the penguin effects described by
$a e^{i\theta}$, we may use a $\bar b\to \bar d c \bar c$ transition, 
as this parameter is here not doubly Cabibbo suppressed \cite{RF-BpsiK,RF-ang}.
In the following, we will use the decay $B^0\to J/\psi \pi^0$.
In Ref.~\cite{CPS}, a similar ansatz was used to constrain the
penguin effects in the golden mode. However, the quality of the data has 
improved such that we go beyond this paper by allowing for 
$\phi_d^{\rm NP}\not=0^\circ$. Moreover, as we will see below, the current 
$B$-factory data point already towards a 
{\it negative} value of $\Delta\phi_d$, where mixing-induced CP violation
in $B^0\to J/\psi\pi^0$ is the driving force, thereby reducing the tension 
(\ref{2b-diff}) in the fit of the UT.

In the SM, we have
\begin{equation}\label{ampl-d}
\sqrt{2}A(B^0\to J/\psi \pi^0)=\lambda{\cal A}'\left[1-a' e^{i\theta'}e^{i\gamma}\right],
\end{equation}
where the  $\sqrt{2}$ factor  is associated with the $\pi^0$ wavefunction,
while ${\cal A}'$ and $a'e^{i\theta'}$ are the counterparts of (\ref{Ampl-def}) 
and (\ref{a-def}), respectively. We see now explicitly that -- in contrast 
to (\ref{ampl}) -- the latter quantity does not enter (\ref{ampl-d}) 
with the $\epsilon$. 
The CP asymmetry $A_{\rm CP}(t;J/\psi\pi^0)$
(see (\ref{CP-asym})) was recently measured by the BaBar (SLAC) 
\cite{BaBar-psipi0} and Belle (KEK) \cite{Belle-psipi0} collaborations, yielding
the following averages \cite{HFAG}:
\begin{eqnarray}
C(J/\psi \pi^0)&=& -0.10 \pm 0.13, \\
S(J/\psi \pi^0)&=&-0.93\pm0.15\,.
\end{eqnarray}
Note that the error of $S(J/\psi\pi^0)$ is that of the HFAG, which is not inflated due
to the inconsistency of the data.

The values of these CP asymmetries allow us to calculate $a'$ as functions
of $\theta'$. We obtain two relations from $C(J/\psi \pi^0)$ and $S(J/\psi \pi^0)$
(${\cal O}=C$ and $S$, respectively),
\begin{equation}\label{a-theta-form}
a'=U_{\cal O}\pm\sqrt{U_{\cal O}^2-V_{\cal O}},
\end{equation}
where
\begin{equation}
U_{C}\equiv\cos\theta'\cos\gamma+\frac{\sin\theta'\sin\gamma}{C(J/\psi \pi^0)},
\quad
V_{C}\equiv1,
\end{equation}
and
\begin{equation}
U_{S}\equiv\left[\frac{\sin(\phi_d+\gamma)+
S(J/\psi \pi^0)\cos\gamma}{\sin(\phi_d+2\gamma)
+S(J/\psi \pi^0)}\right]\cos\theta'
\end{equation}
\begin{equation}\label{VS}
V_S\equiv\frac{\sin\phi_d+S(J/\psi \pi^0)}{\sin(\phi_d+2\gamma)+S(J/\psi \pi^0)}.
\end{equation}
The intersection of the $C(J/\psi \pi^0)$ and $S(J/\psi \pi^0)$ contours 
fixes then the hadronic parameters $a'$ and $\theta'$ in the SM; when 
allowing for an additional NP phase, one has to take into account 
$S(J/\psi K^0)$ together with $S(J/\psi \pi^0)$ in order to have a constraint 
in the $a'$--$\theta'$ plane. From $C(J/\psi K^0)$ comes another 
constraint, which is of the form (\ref{a-theta-form}) with the replacements 
$a'\to \epsilon a$ and $\theta'\to 180^\circ+\theta$. It should be stressed that 
(\ref{a-theta-form})--(\ref{VS}) are valid exactly as these expressions follow 
from the SM structure of $B^0\to J/\psi \pi^0$. 

Neglecting penguin annihilation and exchange topologies, which 
contribute to $B^0\to J/\psi\pi^0$ but have no counterpart  in $B^0\to J/\psi K^0$ 
and are expected to play a minor r\^ole (which can be probed through
$B^0_s\to J/\psi \pi^0$), 
we obtain in the  $SU(3)$ limit 
\begin{equation}\label{a-theta-rel}
a'=a, \quad \theta'=\theta\,.
\end{equation}
Thanks to these relations, we can determine the shift $\Delta\phi_d$ by 
means of (\ref{S-gen})--(\ref{C-psiK}) from the data. We expect them to hold 
to a reasonable accuracy; however, one has to keep 
in mind that sizable non-factorizable effects may induce $SU(3)$-breaking corrections. 
Their impact on the determination of $\Delta\phi_d$ can be easily inferred from 
(\ref{tDelPhi}). Neglecting terms of order $\epsilon^2$, we have a linear dependence 
on $a\cos\theta$. Consequently, corrections to the left-hand side of (\ref{a-theta-rel})
propagate linearly, while $SU(3)$-breaking effects in the strong phases will
generally lead to an asymmetric uncertainty for $\Delta\phi_d$.

Before having a closer look at the picture emerging from the current $B$-factory 
data, let us discuss another constraint which follows from the CP-averaged 
branching ratios. To this end, we introduce
\begin{eqnarray}
H &\equiv &\frac{2}{\epsilon}
\left[\frac{\mbox{BR}(B_d\to J/\psi\pi^0)}{\mbox{BR}(B_d\to J/\psi K^0)}
\right]\left|\frac{{\cal A}}{{\cal A}'}\right|^2
\frac{\Phi_{J/\psi K^0}}{\Phi_{J/\psi\pi^0}}\nonumber\\
&=&\frac{1-2a'\cos\theta'\cos\gamma+a'^2}{1+2\epsilon a\cos\theta\cos\gamma+\epsilon^2a^2},\label{H-def}
\end{eqnarray}
where the $\Phi_{J/\psi P}\equiv\Phi(M_{J/\psi}/M_{B^0},M_{P}/M_{B^0})$ 
are phase-space factors  \cite{RF-BpsiK}.
In order to extract $H$ from the data, we have to analyze 
the $SU(3)$-breaking corrections to $|{\cal A}/{\cal A}'|$. We assume them to be factorizable, and 
thus given by the ratio of two form factors, evaluated at $q^2 = M_{J/\psi}^2$. 
This ratio has been studied in detail using QCD light-cone 
sum rules (LCSR) \cite{Ball:2004ye}. 
We shall use the latest result for the form factor ratio at $q^2 = 0$  
\cite{Duplancic:2008ix,Duplancic:2008tk}, 
\begin{equation}
f^+_{B \to K} (0) / f^+_{B \to \pi} (0) = 1.38^{+0.11}_{-0.10}, 
\end{equation}
and perform the extrapolation to $q^2 = M_{J/\psi}^2$ by using a simple BK 
parametrization
\cite{Becirevic:1999kt}
\begin{equation} \label{BK}
f^+(q^2) = f^+ (0)\left[ \frac{M_B^2 \, M_*^2}{(M_*^2 - q^2)(M_B^2 - \alpha q^2)}\right]. 
\end{equation}
Here $M_*$ is the mass of the ground state vector meson in the relevant channel 
and  the pole at $M^2/\alpha$ models the contribution of the hadronic 
continuum for $q^2 > M_*^2$.
The BK parameter $\alpha$ has been fitted to the $B \to \pi$ lattice 
data to be $\alpha_\pi = 0.53 \pm 0.06$. Nothing is known about the value of 
$\alpha$ for the $B \to K$ form factor and we shall use the  simple assumption 
that the main $SU(3)$-breaking effect is due to the shift of the continuous part 
of the spectral  function from the $B \pi$ to the $B K$ threshold. This leads to 
$
\alpha_K   = 0.49 \pm 0.05 \, , 
$
and -- extrapolating in this way to $q^2 = M_{J/\psi}^2$ --  we get  
\begin{equation}
f^+_{B \to K} (M_{J/\psi}^2) / f^+_{B \to \pi} (M_{J\psi}^2) = 1.34\pm 0.12.
\end{equation}
Using 
$\mbox{BR}(B^0\to J/\psi K^0)=(8.63\pm0.35)\times10^{-4}$ and
$\mbox{BR}(B^0\to J/\psi \pi^0)=(0.20\pm0.02)\times10^{-4}$ \cite{HFAG},
we obtain $H=1.53 \pm0.16_{\rm{BR}}\pm 0.27_{\rm{FF}}$,
where we give the errors induced by the branching ratios and the 
form-factor ratio.  

Using (\ref{a-theta-rel}), we obtain the following relation \cite{RF-BpsiK}:
\begin{equation}
C(J/\psi K^0)=-\epsilon H C(J/\psi\pi^0),
\end{equation}
which would offer an interesting probe for $SU(3)$ breaking. However, 
the value of $H$ given above yields $C(J/\psi K^0)=0.01\pm0.01$, which is 
consistent with (\ref{C-psiK}), but obviously too small for a powerful test.
 
\begin{figure}
    \centering
   \includegraphics[width=5.5truecm]{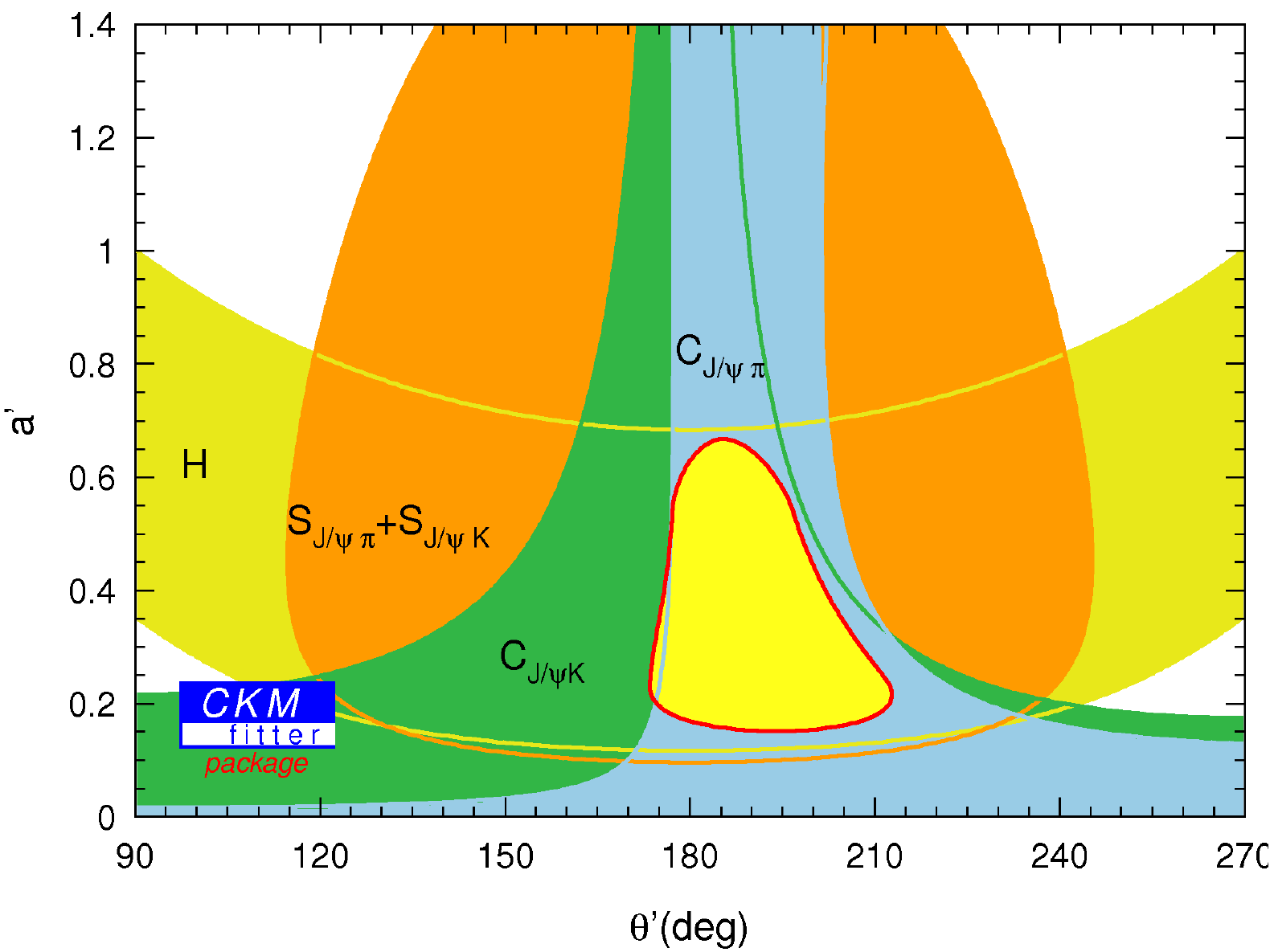} 
      \vspace*{-0.3truecm}
    \caption{The $1\,\sigma$ ranges in the $\theta'$--$a'$ plane 
             with current data.}\label{fig:2}
 \end{figure}
 
If we apply once more (\ref{a-theta-form}) with 
\begin{eqnarray}
U_{H}&=&\left(\frac{1+\epsilon H}{1-\epsilon^2 H}\right)\cos\theta'\cos\gamma \\ 
V_{H}&=&(1-H)/(1-\epsilon^2 H),
\end{eqnarray}
i.e.\ ${\cal O}=H$, we may again calculate $a'$ as function of $\theta'$. 
In contrast to the CP asymmetries of $B^0\to J/\psi\pi^0$, we have to deal 
here with $SU(3)$-breaking effects, which enter implicitly through the 
determination of $H$.

In Fig.~\ref{fig:2}, we show the fits in the $\theta'$--$a'$ plane for the
current data with $1\,\sigma$ ranges. The major implication of 
$S(J/\psi\pi^0)$ is $\theta'\in[90^\circ,270^\circ]$. Looking at 
(\ref{a-def}), this is actually what we expect. $S(J/\psi K^0)$ fixes the
NP phase essentially to $(\phi_d)_{J/\psi K^0}-2\beta_{\rm true}$, as 
the NP phase is an $\mathcal{O}(1)$ effect in $S(J/\psi K^0)$, while the additional 
SM contribution is suppressed by $\epsilon$. The negative central value of 
$C(J/\psi \pi^0)$ prefers $\theta'>180^\circ$. The intersection of the $C(J/\psi\pi^0)$ 
and $H$ bands, which falls well into the $S(J/\psi\pi^0,J/\psi K^0)$ as well as the 
$C(J/\psi K^0)$ region, gives then $a'\in[0.15,0.67]$ and $\theta'\in[174,213]^\circ$ 
at the $1\,\sigma$ level. Note that all three constraints give finally an 
unambiguous solution for these parameters.

\begin{figure}
    \centering
   \includegraphics[width=5.5truecm]{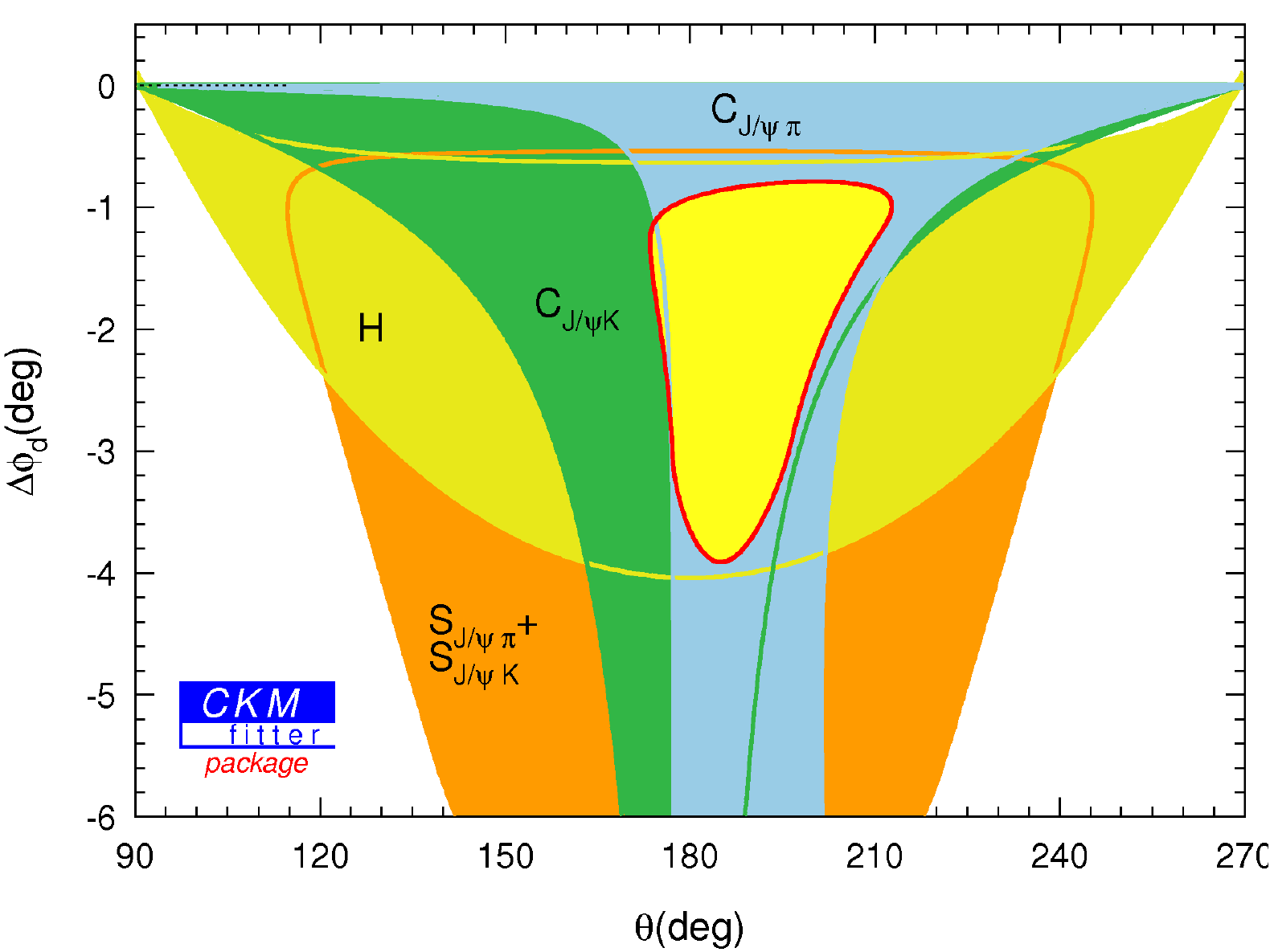}
   \vspace*{-0.3truecm}
     \caption{$\Delta\phi_d$ for the constraints shown in Fig.~\ref{fig:2}.}\label{fig:3}
 \end{figure}

\begin{figure}
    \centering
      \includegraphics[width=5.5truecm]{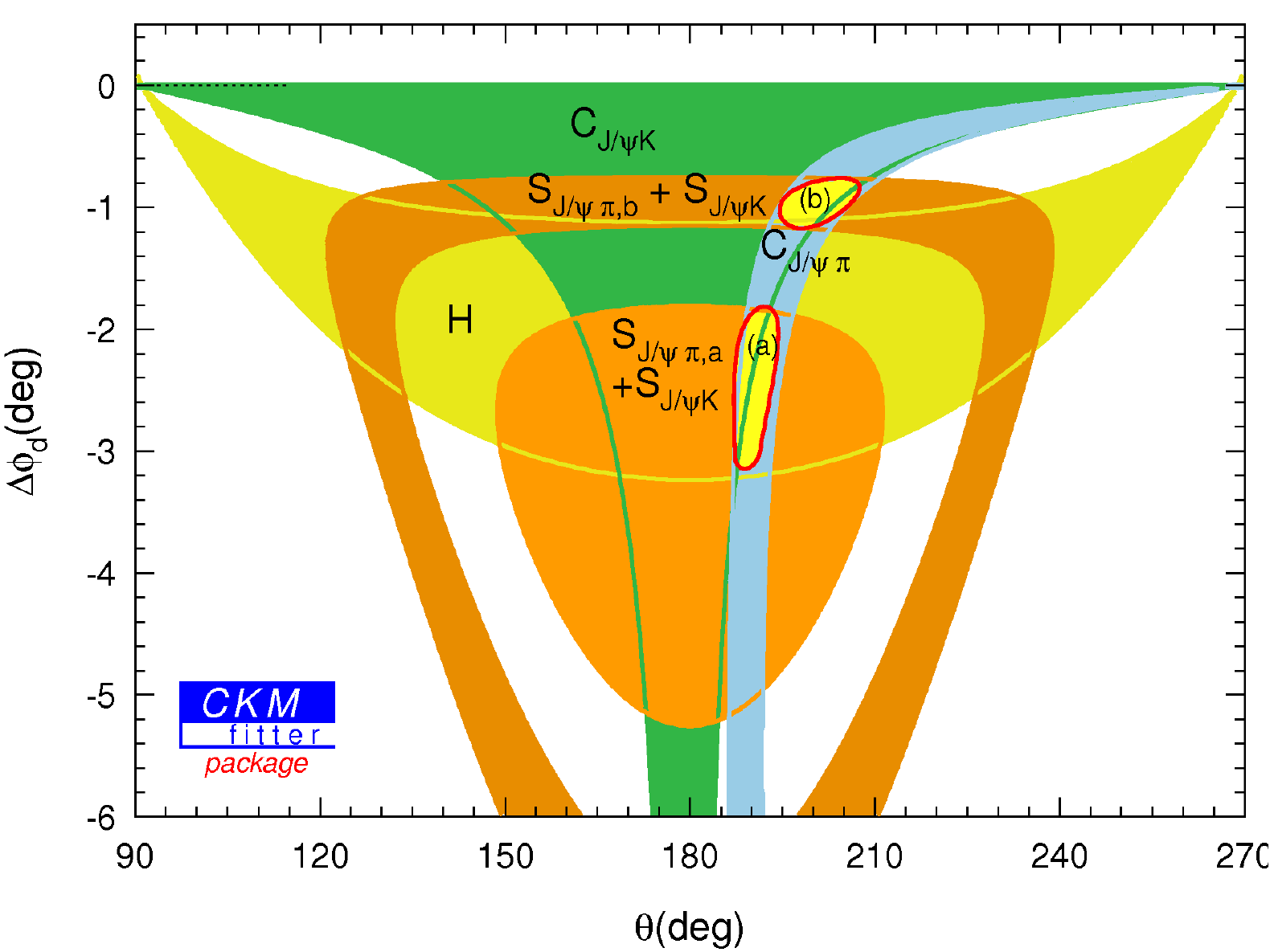}
   \vspace*{-0.3truecm}
     \caption{Future benchmark scenarios, as discussed in the text.}\label{fig:4}
 \end{figure}

In Fig.~\ref{fig:3}, we convert the curves in Fig.~\ref{fig:2} into the
$\theta$--$\Delta\phi_d$ plane with the help of (\ref{a-theta-rel}) and 
(\ref{sDelPhi})--(\ref{tDelPhi}). We see that a negative value of
$\Delta\phi_d$ emerges;  the global fit to all observables
yields $\Delta\phi_d\in[-3.9,-0.8]^\circ$, mainly due to the constraints from 
$H$ and $C(J/\psi\pi^0)$, corresponding to $\phi_d=(42.4^{+3.4}_{-1.7})^\circ$. 
Furthermore, the fit gives $\phi_{d}^{\rm NP}\in[-13.8,1.1]^\circ$, which includes 
the SM value $\phi_d^{\rm NP}=0^\circ$. Consequently,  the negative sign of 
the SM correction $\Delta\phi_d$ softens  the tension in the fit of the UT.

We have studied the impact of $SU(3)$-breaking corrections by setting
$a=\xi a'$ in (\ref{a-theta-rel}) and uncorrelating $\theta$ and $\theta'$. 
Even when allowing for $\xi\in[0.5,1.5]$ and
$\theta,\theta'\in[90,270]^\circ$ in the fit, 
and using a $50\%$ increased error for the form-factor ratio in view of
non-factorizable contributions to $|{\cal A}/{\cal A}'|$, the global fit yields
$\Delta\phi_d\in[-6.7,0.0]^\circ$ and $\phi_d^{\rm NP}\in[-14.9,4.0]^\circ$, determined
now mostly by $C(J/\psi K^0)$ and $H$. Consequently, these 
$SU(3)$-breaking effects do not alter our conclusions.

The increasing experimental precision will further constrain the 
hadronic parameters. However, the final reach for a 
NP contribution to the $B^0_d$--$\bar B^0_d$ mixing phase will strongly depend 
on the measured values of the CP asymmetries of $B^0\to J/\psi\pi^0$, which 
are challenging for LHCb because of the neutral pions
(here a similar analysis could be performed with $B^0_s\to J/\psi K_{\rm S}$ 
\cite{RF-BpsiK}), but can be measured at future super-$B$ factories. 

We illustrate this through two benchmark scenarios, assuming a reduction
of the experimental uncertainties of the CP asymmetries of $B^0\to J/\psi K^0$ 
by a factor of 2, and errors of the branching ratios and $\gamma$ that are 
five-times smaller; the scenarios agree 
in $C(J/\psi \pi^0) = -0.10 \pm 0.03$, but differ in
$S(J/\psi \pi^0)$. In the high-$S$ scenario (a), we assume $S = -0.98 \pm 0.03$.
As can be seen in Fig.~\ref{fig:4}, $\Delta\phi_d\in[-3.1,-1.8]^\circ$ 
(with $a'\sim~0.42$, $\theta'\sim191^\circ$) 
will then come from the lower value of $S$ and $H$, which we assume 
as $H=1.53\pm0.03\pm0.27$. In the low-$S$ scenario (b), we assume 
$S = -0.85 \pm 0.03$. In this case, $\Delta\phi_d\in[-1.2,-0.8]^\circ$ (with 
$a'\sim0.18$, $\theta'\sim201^\circ$) would be determined by $S$ and $C$ 
alone, while $H$ would only be used to rule out the second solution.
By the time the accuracies of these benchmark
scenarios can be achived, we will also have a much better picture of $SU(3)$-breaking
effects through data about $B_{s,d,u}$ decays. 

Since the experimental uncertainty of $(\phi_d)_{J/\psi K^0}$ could be reduced 
to $\sim 0.3^\circ$ at an upgrade of LHCb and an $e^+e^-$ super-$B$ factory, 
these corrections will be essential. It is interesting to note that the quality of 
the data will soon reach a level in the era of precision flavour physics where 
subleading effects, i.e.\ doubly Cabibbo-suppressed penguin contributions, 
have to be taken into account. In particular, in the analyses of CP violation in 
the golden $B^0\to J/\psi K_{\rm S,L}$ modes this is mandatory in order to
fully exploit the physics potential for NP searches.

\paragraph{Acknowledgements:} \noindent
S.F., M.J. and T.M. acknowledge discussions with Th. Feldmann and
the support by the German Ministry of Research (BMBF,
Contract No.~05HT6PSA).


\begin{thebibliography}{99}
%
%
%
\bibitem{bisa}A.~B.~Carter and A.~I.~Sanda,
  Phys.\ Rev.\ Lett.\  {\bf 45}, 952 (1980),
  Phys.\ Rev.\  D {\bf 23}, 1567 (1981);
I.~I.~Y.~Bigi and A.~I.~Sanda,
  Nucl.\ Phys.\  B {\bf 193}, 85 (1981).

\bibitem{BellePsiK}K. -F. Chen {\it et al.}\ [Belle Collaboration],
  Phys.\ Rev.\ Lett.\ {\bf 98}, 031802 (2007).

\bibitem{BaBarPsiK}B. Aubert {\it et al.}\ [BaBar Collaboration],
  arXiv:0808.1903 [hep-ex].

\bibitem{HFAG}E. Barberio {\it et al.}\ [Heavy Flavour Averaging Group], 
  arXiv:0704.3575 [hep-ex];
for the most recent updates, see  http://www.slac.stanford.edu/xorg/hfag. 

\bibitem{CKMfitterSoftware}A.~H\"ocker, H.~Lacker, S.~Laplace and  F.~Le Diberder,
  Eur.\ Phys.\ J. C {\bf 21}, 225 (2001).

\bibitem{wolf} L.~Wolfenstein,
  Phys.\ Rev.\ Lett.\  {\bf 51}, 1945 (1983).
  
\bibitem{blo} A.~J.~Buras, M.~E.~Lautenbacher and G.~Ostermaier,
  Phys.\ Rev.\  D {\bf 50}, 3433 (1994).

\bibitem{PDG}Particle Data Group (2008), see
http://pdg.lbl.gov/.

\bibitem{UTfit}M.~Bona {\it et al.}~[UTfit Collaboration],
  JHEP {\bf 0507}, 028 (2005); updates: http://utfit.roma1.infn.it/.

\bibitem{CKMfitter}J.~Charles {\it et al.}~[CKMfitter Group], 
Eur.\ Phys.\ J.\ C {\bf 41}, 1 (2005); updates: http://ckmfitter.in2p3.fr/.

\bibitem{RF-BsKK-07}R.~Fleischer,
  Eur.\ Phys.\ J.\  C {\bf 52}, 267 (2007).
  
\bibitem{FM-BpsiK}R.~Fleischer and T.~Mannel,
  Phys.\ Lett.\  B {\bf 506}, 311 (2001).
  
\bibitem{UTfit-NP}M.~Bona {\it et al.}  [UTfit Collaboration],
  JHEP {\bf 0603}, 080 (2006).

\bibitem{BaFl}P.~Ball and R.~Fleischer,
  Eur.\ Phys.\ J.\  C {\bf 48}, 413 (2006).

\bibitem{RF-BpsiK} R.~Fleischer,
  Eur.\ Phys.\ J.\  C {\bf 10}, 299 (1999).

\bibitem{FFM-long}S. Faller, R. Fleischer and T. Mannel, CERN-PH-TH/2008-167.

\bibitem{RF-ang}R.~Fleischer,
  Phys.\ Rev.\  D {\bf 60}, 073008 (1999).

\bibitem{CPS}M.~Ciuchini, M.~Pierini and L.~Silvestrini,
  Phys.\ Rev.\ Lett.\  {\bf 95}, 221804 (2005).

\bibitem{BaBar-psipi0}B.~Aubert {\it et al.}  [BaBar Collaboration],
  arXiv:0804.0896 [hep-ex].

\bibitem{Belle-psipi0} S.~E.~Lee {\it et al.}\  [Belle Collaboration],
  Phys.\ Rev.\  D {\bf 77}, 071101 (2008).

\bibitem{Ball:2004ye}
  P.~Ball and R.~Zwicky,
  Phys.\ Rev.\  D {\bf 71}, 014015 (2005).
 
\bibitem{Duplancic:2008ix}
  G.~Duplancic, A.~Khodjamirian, T.~Mannel, B.~Melic and N.~Offen,
  JHEP {\bf 0804}, 014 (2008).
  
\bibitem{Duplancic:2008tk}
  G.~Duplancic and B.~Melic,
  arXiv:0805.4170 [hep-ph].
  
\bibitem{Becirevic:1999kt}
  D.~Becirevic and A.~B.~Kaidalov,
  Phys.\ Lett.\  B {\bf 478}, 417 (2000).

%
%
%
\end{thebibliography}
\end{document}